# Anvendelse av kunstig intelligens (KI) i Norge i norsk offentlig sektor 2024

John Krogstie

Norwegian University of Science and Technology (NTNU) {John.Krogstie@ntnu.no}

**Sammendrag.** Det er store forventninger til bruk av KI i Norge. På den annen side rapporteres det at adopsjonen av KI i Norge går tregere enn forventet både i privat og offentlig sektor. Ved hjelp av svar fra NOKIOS teknologiradar 2017-2021, IT i Praksis – undersøkelser utført av Rambøll i 2021-2024, samt en annen nasjonal undersøkelse som en del av en femårig syklus, ser vi i denne artikkelen på rapportert og planlagt bruk av KI med fokus på lokale (kommuner) og nasjonale offentlige etater. IT i praksis distribueres til en lang rekke norske offentlige virksomheter, med en svarprosent på over 50%. De nyeste dataene (2024) presentert i denne artikkelen er basert på svar fra 335 offentlige organisasjoner, med 237 kommuner, og 98 offentlige organisasjoner på nasjonalt eller regionalt nivå. Undersøkelsen begrefter at bruken av KI fortsatt er på et tidlig stadium, selv om forventningene er høye til fremtidig bruk

**Keywords:** KI, implementasjon, spørreundersøkelser

## 1  Introduksjon

Digitaliseringen av samfunnet har pågått gradvis i minst de siste 40 årene (selv om man kan spore systemer tilbake til 60-tallet [6]). De siste årene har digitaliseringen tilsynelatende skutt fart, og fokus siden høsten 2022 har vært på kunstig intelligens. I 2024 fikk vi i Norge et departement for digitalisering og forvaltning (egentlig ikke så ulikt ansvarsområdet som fornyings og administrasjon-departementet (FAD) hadde for snaut 20 år siden), og ministeren der, Karianne Tung, har blant annet blitt kjent for å stille en forventning om at 80% av norske offentlige virksomheter skal bruke KI i 2025.

Potensialet og behovet for å ta i bruk kunstig intelligens i offentlig sektor er tatt opp i en rekke stortingsmeldinger, blant annet Meld. St. 27 (2015–2016) Digital agenda for Norge – IKT for en enklere hverdag og økt produktivitet [15] og Meld. St. 30 (2019–2020) En innovativ offentlig sektor — Kultur, ledelse og kompetanse [18], og i de to statlige strategiene – Nasjonal strategi for kunstig intelligens [17] og Én digital offentlig sektor. Digitaliseringsstrategi for offentlig sektor 2019–2025 [16]. Stortingsmeldingene og strategidokumentene legger vekt på at offentlig sektor skal (og må) effektiviseres gjennom digitalisering.

På den annen side er det mer som må på plass for å verdi av en teknologi som KI, utover selve teknologien [3]. Nylige rapporter gir en lite positivt bilde av situasjonen.

Riksrevisjonen skrev nylig (Bruk av kunstig intelligens i staten Dokument 3:18 (2023−2024)) [19].
«
- *Statlige virksomheter utnytter mulighetene med kunstig intelligens ulikt, og kunstig intelligens er fortsatt lite i bruk.*
- *Viktige forutsetninger for å ta i bruk kunstig intelligens i større skala er fortsatt ikke på plass.*
- *Samordningen av arbeidet med kunstig intelligens i offentlig sektor er mangelfull, og den samlede innsatsen er for svak gitt ambisjonen om at Norge skal ha en infrastruktur for kunstig intelligens i verdensklasse.* «

Det står ikke spesielt mye bedre til i privat sektor. En undersøkelse fra Tekna [21] blant deres medlemmer (som kan være både i privat og offentlig sektor, men de har flest medlemmer i privat sektor) viser at kunstig intelligens i liten grad er tatt aktivt i bruk i virksomhetene hvor medlemmene er ansatt. Kun 7 prosent av medlemmene svarer at virksomheten de er ansatt i, i svært stor eller stor grad (4-5 på en 1-5 skala) benytter seg av KI-verktøy. 4 prosent har KI som en integrert del av virksomhetens strategi og drift og 3 prosent svarer at det er omfattende bruk av KI i alle avdelinger og for de fleste oppgaver. 68 prosent jobber i virksomheter der man i liten eller begrenset grad bruker KI, Undersøkelsen ble sendt ut til et tilfeldig utvalg på 10 000 av Teknas yrkesaktive medlemmer i juni 2024. 13 prosent svarte på undersøkelsen, i alt 1103 personer. 24 prosent av svarene er fra medlemmer med IT bakgrunn.

Et tilsvarende bilde tegnes i Abelias omstillingsbarometer [1], der det fremgår at bare hver 3. virksomhet med 10 personer eller mer har brukt minst en KI-teknologi (som er lavere enn mange av de land vi vanligvis sammenligner oss med i slike undersøkelser). Mens disse undersøkelsen har samlet data, har vi ikke direkte tilgang til disse, så det er vanskelig å gjøre ytterligere analyser. Vi har derimot tilgang til data fra NOKIOS teknologiradar 2017-2021, IT i praksis 2021-2024, samt nasjonale, femårig undersøkelser gjort fra 1993 (sist gang i 2023), der det den siste også har konkrete spørsmål knyttet til anvendelse av KI i sentrale applikasjoner.

I seksjon 2 gir vi mer bakgrunn på utviklingen av digitaliseringen i Norge generelt, og bruk av KI spesielt. Seksjon 3 gir mer bakgrunn for metoden brukt i sammenheng med spørreundersøkelsene vi presenterer resultater fra her. Seksjon 4 presenterer resultater, med fokus på IT i praksis 2021-2024, mens disse diskuteres i lys av bakteppet her i introduksjonen og bakgrunn i seksjon 5 der vi også konkluderer artikkelen og skisserer videre arbeid.

## 2    Bakgrunn

Bruk av data, EDB, IKT og nå mest kjent som digitalisering har pågått i lengre tid. Siden 1993 har nasjonale undersøkelser studert ressursbruk innen feltet i fem-årlige undersøkelser. I [4] oppsummeres blant annet fordeling av tid på ulike arbeidsoppgaver som gjengitt i Figur 1. (med basis i tall opprinnelig publisert etter hver undersøkelse)

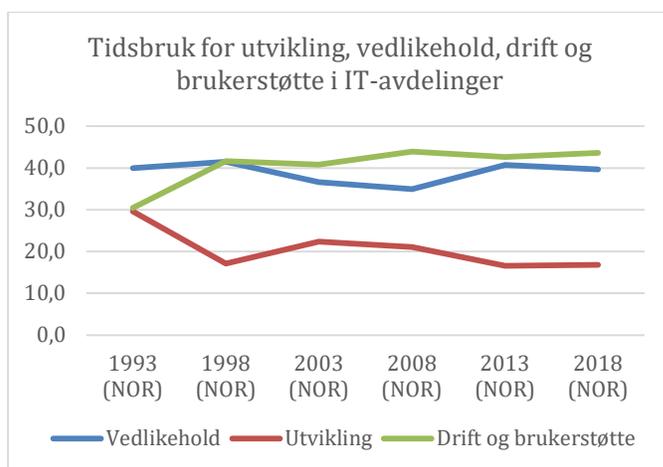

**Fig. 1.** Fordeling av tid på IT-aktivitet i norske virksomheter

Noe som slår en fra Figur 1 er at etter årtusen-skiftet er fordelingen av tidsbruk i gjennomsnitt svært stabil selv om teknologien vi bruker (for eksempel sky-løsninger og bruk av KI) og metode (med mer og mer fokus på smidighet, produktorganisering og devops) har endret seg mye. Et annet område som er stabilt, er andel av nye systemer som er erstatningssystemer, det vil si nye systemer som erstatter eksisterende systemer uten å gi mye ny funksjonalitet (i hvert fall i første leveranse). Helt siden 1993 har mer enn halvparten av 'nye' IT-systemer laget i norske virksomheter vært erstatningssystemer.

Både i riksrevisjonens rapport og i IT i praksis brukes følgende definisjon av KI (hentet fra KI-strategien for Norge laget i 2020 [17]])

Den nasjonale strategien for kunstig intelligens beskriver kunstig intelligens som følger: KI-systemer er

> «*systemer som utfører handlinger, fysisk eller digitalt, basert på tolkning og behandling av strukturerte eller ustrukturerte data, i den hensikt å oppnå et gitt mål. Enkelte KI-systemer kan også tilpasse seg gjennom å analysere og ta hensyn til hvordan tidligere handlinger har påvirket omgivelsene*".

IT i praksis eksemplifiserer ytterligere for å gjøre det enklere å være enig for respondenter om de har å gjøre med KI-systemer:

> "*Kunstig intelligens brukes i dag i mange ulike applikasjoner, som for eksempel chatbots, bildesøk, personlige assistenter, talegjenkjenning, anbefalingssystemer og selvkjørende biler.*"

Dette tilsvarer definisjonen som ble brukt i NOKIOS teknologiradar 2017-2021 [12]. NOKIOS teknologiradar var spørreundersøkelser blant norske virksomheter (både offentlig og private) knyttet til bruk av ny teknologi laget i sammenheng med den årlige nasjonale NOKIOS-konferansen (om IKT i offentlig sektor), og presentert der og i rapporter på nett. Figurene 2 til 5 viser viktig tilbakemeldinger knyttet til modenhet av KI, modenhet av egen virksomhet til å ta i bruk KI, samt status for anvendelse av KI i virksomheten.

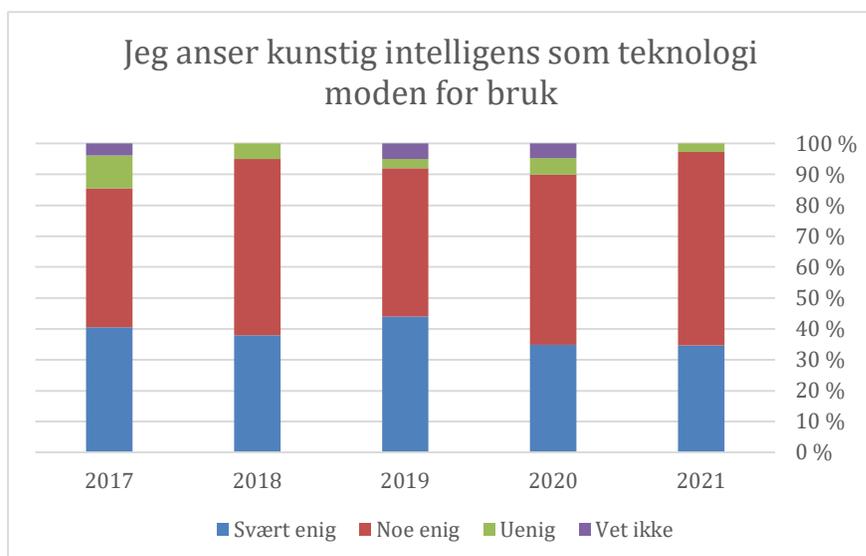

**Fig. 2.** Anslag av modenhet av kunstig intelligens - Teknologiradaren 2017-2021

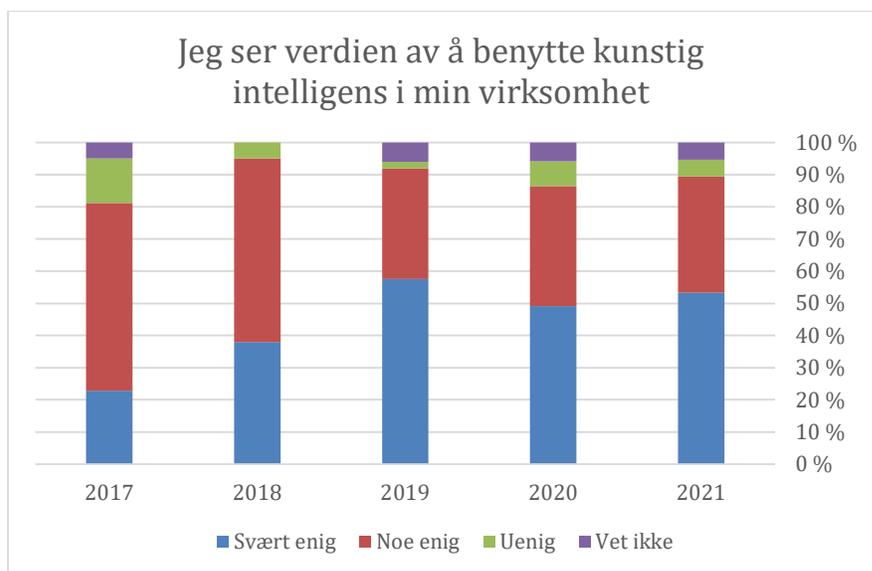

**Fig. 3.** Anslag av verdi av KI i norske virksomheter - Teknologiradaren 2017-2021

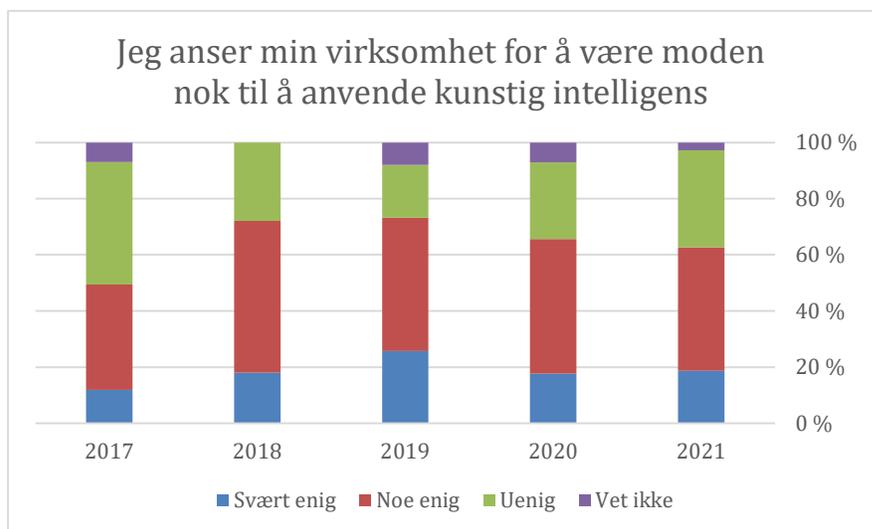

**Fig. 4.** Anslag modenhet av norske virksomheter i å ta i bruk KI - Teknologiradaren 2017 - 2021

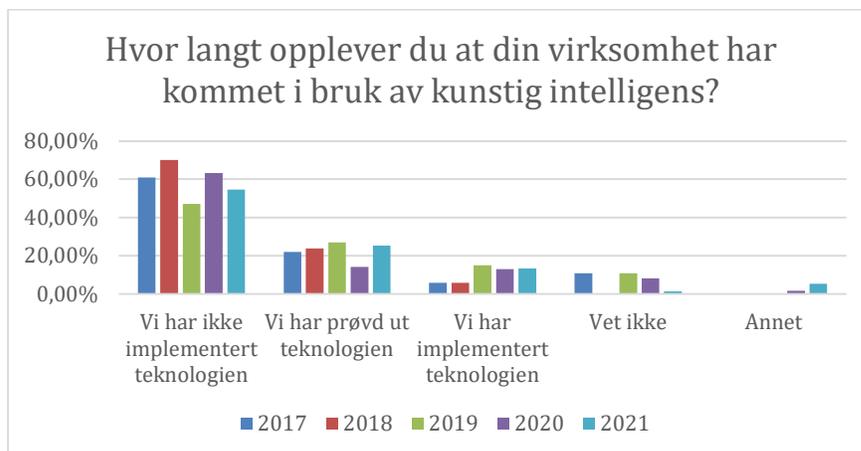

**Fig. 5.** Status for bruk av KI i norske virksomheter - Teknologiradaren 2017 -2021

Den største utfordringen for å ta i bruk KI som nevnes er mangel på kompetanse. Da det var liten utvikling i bruk av ny teknologi på denne tiden, ble teknologiradaren ikke fulgt opp videre, men en del av problemstillingene tas videre i undersøkelsene vi presentere senere i artikkelen.

## 3   Metode

Vi vil i artikkelen ha basis i flere spørreskjema-undersøkelser. Undersøkelsen «IT i praksis» sendes ut til rundt 500 offentlige virksomheter hvert år. I 2024 ble 335 av de distribuerte spørreskjemaer i «IT i praksis» returnert. Tidligere undersøkelser har også hatt en svarprosent på over 50 %. Dette er en høy svarprosent for slike undersøkelser, men likevel er det begrensninger ved spørreskjemametoder, som vi vil diskutere litt mer detaljert nedenfor. Se [13,14] for mer informasjon om hvordan undersøkelsene gjennomføres.

Den andre nasjonale undersøkelse følger opp en fem-årlig syklus av undersøkelser som er gjort blant norske virksomheter tilbake til 1993. Vi sammenligner sentrale tall med disse for å se at de gir et tilsvarende bilde på områder som vi har sett har vært stabile (dvs. vi vil forvente stabile tall på tidsfordeling mellom utvikling, vedlikehold og andre aktiviteter, samt andel nye systemer som er erstatningssystemer som beskrevet i seksjon 2) og tolker tall knyttet til KI-bruk i lys av dette.

En spørreundersøkelse av denne formen har kjente begrensninger [7, 8]. I vårt tilfelle hadde vi (i IT i praksis) et større antall svar enn i tidligere undersøkelser, og en svarprosent på rundt 50-67 % med svar fra 250-300 organisasjoner for hver undersøkelse gir oss økt tillit til at resultatene gir et realistisk bilde av situasjonen.

De fleste som svarte, ledet IT-aktiviteten i organisasjonen. De kan ha et annet syn på virkeligheten enn IT-utviklere. For eksempel fant Jørgensen [7] at en leder anslår andelen korrigerende vedlikehold til å være for høy når den er basert på antagelser i stedet for gode data, se også [20] som rapporterer en lignende effekt. Alle våre undersøkelser har imidlertid data fra IT-ledere, og det er derfor rimelig å sammenligne disse undersøkelsene når man ser på trender.

For å oppnå konsistente svar kreves det at respondentene har en felles forståelse av de grunnleggende konseptene i undersøkelsesskjemaet. Dette kan være vanskelig å sikre i praksis. Jørgensen [7] fant for eksempel at respondentene brukte sin egen definisjon av «vedlikehold av programvare» selv om begrepet var definert i begynnelsen av spørreskjemaet. 'Bruk av KI' er et annet sentralt konsept som var forsøkt forklart slik at alle skulle tolke det likt. Pilotstudier blir gjennomført hvert år i flere virksomheter for å avdekke uklare spørsmål i IT i praksis for å begrense utfordringer med terminologi.

Blant risikoene ved utforming av spørreskjemaer er ledende eller sensitive spørsmål, noe som kan resulterer i partiske eller uærlige svar. Vi mener at vi stort sett har unngått dette problemet. Vi lovet og iverksatte full konfidensialitet til respondentene.

Et annet problem er at alle undersøkelsene er gjort i Norge. Da de første undersøkelsene ble gjennomført i 1993 [11] ble disse sammenlignet med de viktigste internasjonale undersøkelsene på den tiden, og vi fant lignende mønstre som det som var rapportert i andre land. «IT i praksis» har vært gjennomført i Danmark i mer enn 20 år, og det ville vært interessant å sammenligne resultatene fra de norske studiene med tilsvarende studier gjort i Danmark.

## 4    Oversikt over resultater

KI i ulike former har lenge vært en lovende teknologi, og er en av teknologiområdene som ble undersøkt i NOKIOS teknologiradar allerede 2017-2021 som vi så seksjon 2. I IT i praksis er det en rekke spørsmål knyttet til KI, både bruk nå og forventet bruk de neste tre årene

Spørsmålet 'I hvilken grad har din virksomhet tatt i bruk kunstig intelligens i oppgaver og tjenester?' ble besvart på en 5-punkts skala (1 - I svært liten grad; 2 - I liten grad; 3 - I noen grad; 4 -I stor grad; 5 - I svært stor grad). I Fig.6 ser vi et gjennomsnitt av dette som et første forsøk på å bedømme utviklingen. Denne figuren indikerer en stagnasjon på under 2.

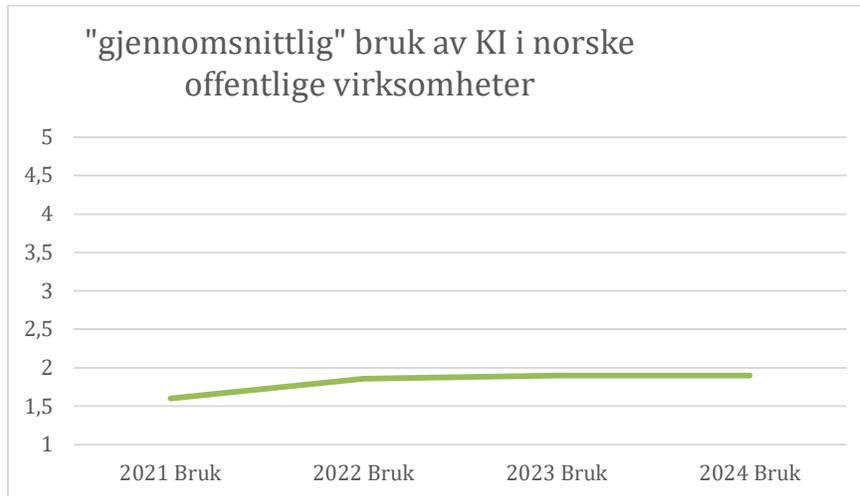
**Fig. 6.** Utvikling på bruk av KI

I figure 7, ser vi dette gjennomsnittet sammenlignet med hvilket nivå man så for seg å ligge på nå (tall fra 2021 som spår om 2024), og hvor man ser for seg å ligge om 3 år innen ulike områder. (har satt tall for 2024 på 1,9 for alle områdene siden det ikke er tall nedbrutt på nåværende område for faktisk bruk.

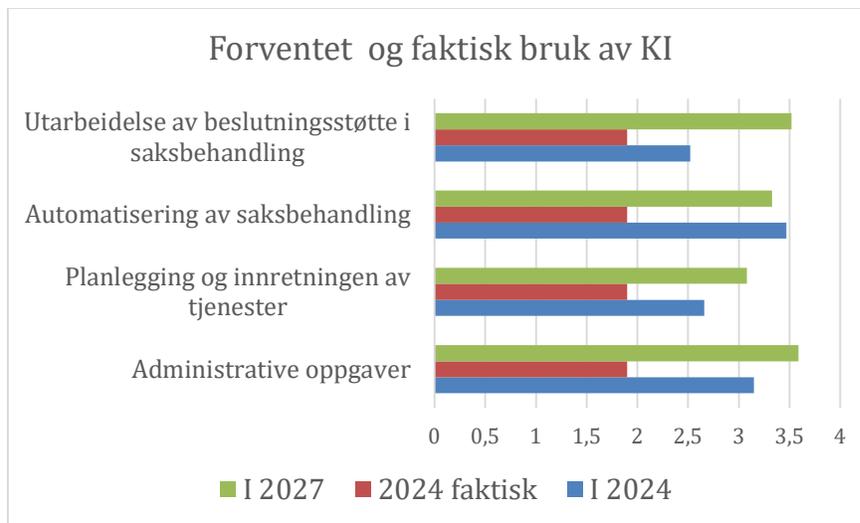
**Fig. 7.** Sammenligning av forventet og faktisk bruk av KI innen ulike områder

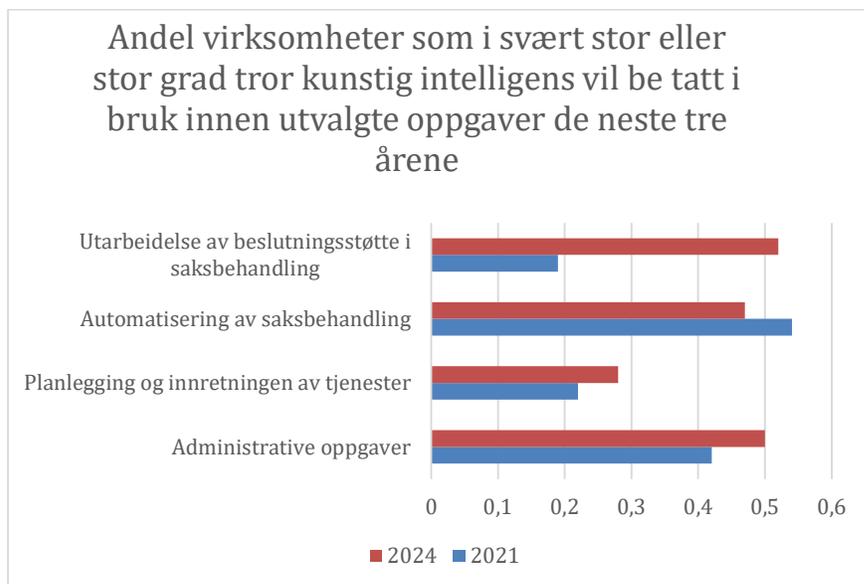

**Fig. 8.** Forventet utvikling av KI-bruk

Figur 8 viser mer om utviklingen av forventning, her de som i stor eller svært stor grad tror KI vil bli brukt til ulike typer oppgaver. Gitt den store fokuset på KI, spesielt etter lansering av store språkmodeller og ChatGPT er det noe overraskende at det ser ut til at bruk har stagnert, mens forventningen også i 2021 om hvor vi ville være i 2024 (om tre år) er svært mye høyere.

Så hvor nær er vi Tung's 80%? Dette er selvfølgelig avhengig av hva man mener med å ta i bruk.

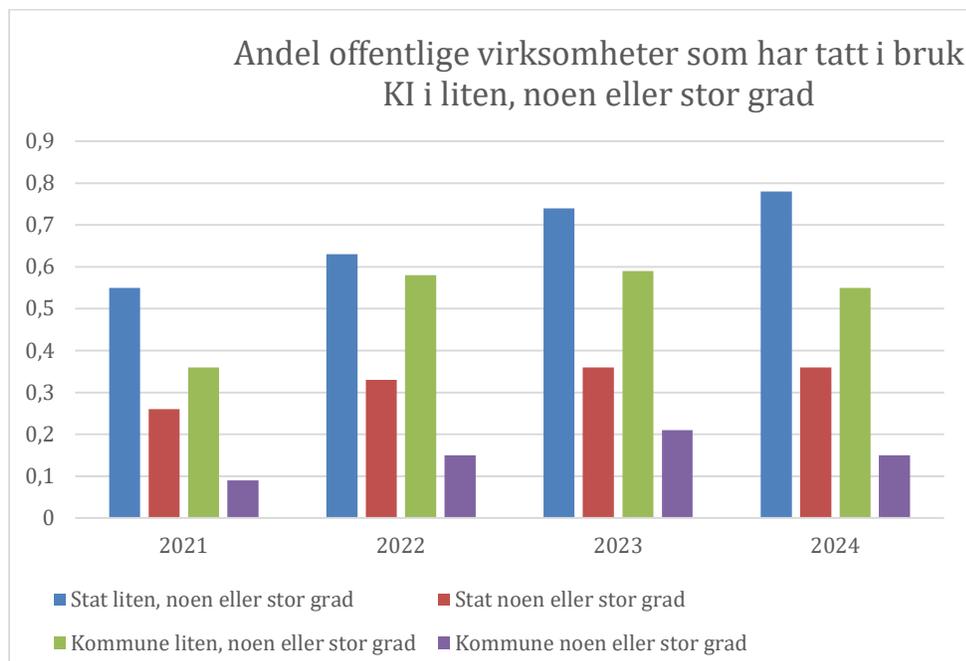

**Fig. 9.** Sammenligning av bruk av KI i statlig og kommunal sektor

I figur 9 ser vi at *om* vi tar med alle som indikerer at de har tatt i bruk KI (selv kun i en liten grad) er vi i statlig sektor nær dette målet. Også etter et slikt beskjedent mål henger kommunal sektor etter. Så hvor butter det?

På spørsmålet 'I hvilken grad opplever virksomheten følgende barrierer for å ta i bruk kunstig intelligens i oppgaver og tjenester (1 - I svært liten grad; 2 - I liten grad; 3 - I noen grad; 4 -I stor grad; 5 - I svært stor grad)' har vi tatt ut andel som svarer 4 eller 5 (I stor eller svært stor grad) (tilsvarende som i [14]). Vi får da følgende rekkefølge på problemområders viktighet (nedbrutt på statlig og kommunal sektor i parentes).

1. 47%: Regulatoriske hindringer / bekymringer (Stat 53%, Kommune 45%)
2. 47%: Manglende innsikt i hvordan KI-verktøy kan løse utfordringer (Stat 25%, Kommune 54%)
3. 46%: For høye investerings- og opplæringskostnader (Stat 31%, Kommune 52%)
4. 45%: Manglende tilgang, format eller kvalitet på data (Stat 41%, Kommune 46%)

5. 42%: Manglende innsikt i tilbudet/markedet av leverandører av KI-verktøy (Stat 18%, Kommune 50%)
6. 39%: Manglende kompatibilitet med eksisterende datasystemer (Stat 22%, Kommune 45%)
7. 33%: Usikkerhet knyttet til gevinster forbundet ved ny teknologi (Stat 20%, Kommune 28%)
8. 24%: Manglende tilbud fra leverandører av KI-verktøy (Stat 8%, Kommune 30%)
9. 18%: Lite endringsvilje i organisasjonen / motstand blant ansatte (Stat 1%, Kommune 24%)

I de 5-årige undersøkelsene [22] (som kun fikk 16 svar i 2023, mot rundt 70 - 80 tidligere) fikk vi også innspill på anvendelse av KI. For å se at vi allikevel kan ha grunn til forvente at overordnede resultater er på tilsvarende nivå som tidligere så vi først på tall for ressursbruk sammenlignet med tidligere undersøkelser (se tabell 1).

Tabell 1. Fordeling av ressursbruk i ulike undersøkelser

| Kategori | 1993 [11] (NOR) | 1998 [5] (NOR) | 2003 [10] (NOR) | 2008 [2] (NOR) | 2013 [9] (NOR) | 2018 [4] (NOR) | 2020 [13] (IT-P) | 2023 [22] (NOR) | 2024 [14] (IT-P) |
|---|---|---|---|---|---|---|---|---|---|
| Vedlikehold | 40,0 | 41,5 | 36,6 | 34,9 | 40,7 | 39,6 | 41,73 | 39,7 | 40,37 |
| Utvikling | 29,6 | 17,1 | 22,4 | 21,1 | 16,6 | 16,8 | 15,8 | 20,7 | 14,01 |
| Drift og brukerstøtte | 30,4 | 41,6 | 40,8 | 43,9 | 42,6 | 43,60 | 41,09 | 39,59 | 46,28 |

Vi ser at den siste undersøkelsen har noe mer utviklingstid enn tidligere undersøkelser. Det vises i [22] at denne forskjellen ikke er signifikant. Vi ser at IT i praksis i 2020 og 2024 ligger omtrent på samme nivå som de øvrige undersøkelsene.

Virksomhetene ble også spurt om hvor langt de har kommet med bruk av KI (som kan sammenlignes med tallene fra teknologiradaren gjengitt i Figur 5).

- Vi har ikke implementert teknologien: 56% (teknologiradaren 2021 55%)
- Vi har prøvd ut teknologien: 18,8% (teknologiradaren 2021 25%)
- Vi har implementert teknologien: 25% (teknologiradar 2021 13%)

Dvs. omtrent like mange sitter på gjerdet som i 2021, mens noen flere av de som prøvde ut teknologi da kan se ut til å ha implementert den. Merk at det ikke er de samme virksomhetene som har besvart de ulike undersøkelsene.

IT i praksis har en litt annen spørsmålsstilling, men om vi gjør en tilsvarende strukturering finner vi:

- Vi har ikke implementert teknologien: 77% (Statlig 47%, kommunalt 88%)
- Vi har prøvd ut teknologien: 19% (Statlig 42%, kommunalt 10%)
- Vi har implementert teknologien: 4% (Statlig 10%, kommunalt 2%)

En ting er å bruke en teknologi, en annen er hvor bredt det brukes. Med basis i spørsmål om eksisterende portefølje av sentrale systemer i virksomheten, systemer under utvikling, og bruk av KI i disse fremgikk det at 2% av eksisterende portefølje har delsystemer som benytter kunstig intelligens. 10% av systemer under utvikling har planlagte del-systemer som benytter kunstig intelligens.

## 5 Diskusjon, konklusjon og videre arbeid

Vi har i denne artikkelen sett på utviklingen av KI i norske virksomheter. Staten har en ambisjon om at 80% av norske offentlige virksomheter skal ta i bruk KI som del av digitaliseringen. Selv om forventingen om hva det vil si å ta i bruk KI er vagt definert, må det sies at vi fortsatt er langt fra reell bruk av KI i offentlig sektor på generell basis. Vi samstemmer uansett med i IT i praksis [14] sitt ønske om at man operasjonaliserer nærmere hva man mener det betyr å bruke en teknologi som KI, selv om mistenker at akkurat dette tallet er satt for å understreke at det skal ses på som helt normalt å bruke denne teknologien (og at det er et unntak å ikke bruke den). Spesielt i kommunal sektor ser det noe overraskende ut som anvendelse av KI har stagnert, selv om fokus på KI etter utbredelse av generativ KI aldri har vært så høyt. Uten at vi har undersøkt dette i detalj, ser det også ut til at privat sektor i Norge tar i bruk KI ganske langsomt, og få virksomheter rapportere at mange av de eksisterende hovedsystemene har en sentral KI-modul.

Det er også interessant å se hvordan forventningene til fremtidig bruk av KI har ligget høyt over faktisk bruk. Vi ser nå at forventningen fortsatt er høy (og økende) selv om det er en mer realistisk forventing om at KI skal støtte snarere enn å automatisere offentlig saksbehandling. Fokus på KI som støttefunksjon, snarere enn som automatisering er trolig også noe som privat sektor må ta høyde for. Selv om krav om transparens er spesielt uttalt i offentlig saksbehandling, vil også nye KI-løsninger i privat sektor forventes å ha en høyere etterrettelighet grunnet behov for å være i henhold til den nye KI-forordningen fra EU.

På kort sikt planlegger vi å samarbeide med Rambøll om «IT i praksis» også i 2025 for å få flere datapunkter som kan bekrefte eller avkrefte mønsteret som finnes i denne undersøkelsen. IT i praksis har på mange måter lagt seg tett opp til målene i digitaliseringsstrategien fra 2019, oppdatert gjennom KI-strategien som kom i januar 2020. Ny digitaliseringsstrategi for Norge ble lansert i september 2024, men vi forventer ikke at KI har en svakere fokus, slik at vi kan følge opp de spørsmålsstillingene man har hatt på dette.

IT i praksis 2024 ble også lansert i september 2024, og vi har bare begynt å analysere data fra denne i mer detalj, så ytterligere resultater vil utarbeides fremover. En langsiktig plan er å gjøre en tilsvarende undersøkelse som gjort med fem-årig mellomrom siden 1993 i 2028 parallelt med å følge opp IT i praksis-undersøkelsene, og i den ha større fokus på å få tilstrekkelig med data som kan gi basis for en mer detaljert statistisk analyse.

I seksjon 3 har vi diskutert noen av de metodiske utfordringene denne type undersøkelser har, som forsterkes ytterligere ved at de nåværende undersøkelsene i kun noen grad er tett koordinert, for eksempel når det gjelder terminologi. Dette gjør det uansett vanskelig å gjøre detaljerte statistiske analyser av data fra IT i praksis, mens vi for egen del har mer kontroll over prosessen, gitt vi greier å få tilstrekkelig antall respondenter.

## Takk

Vi vil takke Rambøll som har gjennomført undersøkelsene IT i praksis, og tillater oss å bruke datamaterialet samt bidra med spørsmål. Vi vil også takke deltakere av spørreundersøkelsen som er gjennomført gjennom årene og våre samarbeidspartner i NOKIOS knyttet til utarbeidelse av teknologiradaren.